\documentclass{ws-procs975x65}

\begin{document}

%to switch ON running title
\markboth{D. Neilsen}{Adaptive Mesh Refinement and Relativistic MHD}

\wstoc{Adaptive Mesh Refinement and Relativistic MHD}{D. Neilsen}

\title{ADAPTIVE MESH REFINEMENT AND RELATIVISTIC MHD}

\author{DAVID NEILSEN, ERIC W.\ HIRSCHMANN}
\address{Department of Physics and Astronomy, Brigham Young University,\\
  Provo, UT 84602, USA}

\author{MATTHEW ANDERSON}
\address{Department of Physics and Astronomy, Louisiana State University,\\
   Baton Rouge, LA 70803-4001, USA}

\author{STEVEN L.\ LIEBLING}
\address{Department of Physics, Long Island University---C.W. Post Campus,\\
  Brookville, NY 11548, USA}

\begin{abstract}

We solve the general relativistic magnetohydrodynamics equations
using distributed parallel adaptive mesh refinement. 
We discuss strong scaling tests of the code,
and present evolutions of Michel accretion and a TOV star.
\end{abstract}

\bodymatter

\section{Introduction}\label{intro}

Compact objects combine a wide array of fascinating physics, and gravitational
waves may open new ways to probe these objects.   We are interested in
systems where gravitational and magnetic fields are dynamically important.
One challenge in simulating astrophysical compact objects 
is that these systems require a range of important
length and time scales.  Adaptive mesh refinement (AMR) thus becomes an
increasingly important tool for large scale computations.  
Furthermore, large computational problems on today's computers must be able 
to effectively utilize a large number of distributed processors.

To address some of the challenges in studying compact objects with numerical
relativity, we have developed a code to solve the general
relativistic magnetohydrodynamics (GRMHD) equations with AMR. 
We use the {\sc had} infrastructure, a modular code for solving
hyperbolic and elliptic differential equations with distributed parallel
AMR.
{\sc had} uses Berger--Oliger style AMR with sub-cycling in time.
Refinement criteria may be problem specific, or a shadow 
hierarchy allows one to easily estimate the truncation error dynamically for 
use in specifying refinement criteria.  The equations to be solved for
a specific problem are isolated in equation
modules, which may be used independently or
combined with other modules.  For example, the MHD and GR equations
are in separate modules, which may be used independently or
combined for the GRMHD code.  

The MHD equations are solved using the Convex Essentially Non-Oscillatory 
(CENO) method, a third order scheme for smooth fluid flow.   
Although our AMR driver can accommodate both finite difference and
finite volume discretization methods, we choose
a finite difference high-resolution shock-capturing
method for the fluid equations to simplify the combined GRMHD code. 
We use hyperbolic divergence cleaning to control the $\nabla \cdot {\bf B}=0$ 
constraint for the magnetic field.  Communication between
coarse and fine grids uses WENO interpolation, a scheme 
designed for discontinuous functions.  Finally, the method of lines is used
for the temporal discretization, and we use a TVD-preserving, 
third-order Runge--Kutta scheme to integrate the equations.
In this paper, we briefly summarize some of our results.
The details of our method and more extensive tests are presented 
elsewhere~\cite{Neilsen2005,Anderson:2006}.

\section{Results} \label{results}

\begin{figure}
\begin{tabular}{lr}
\epsfig{file=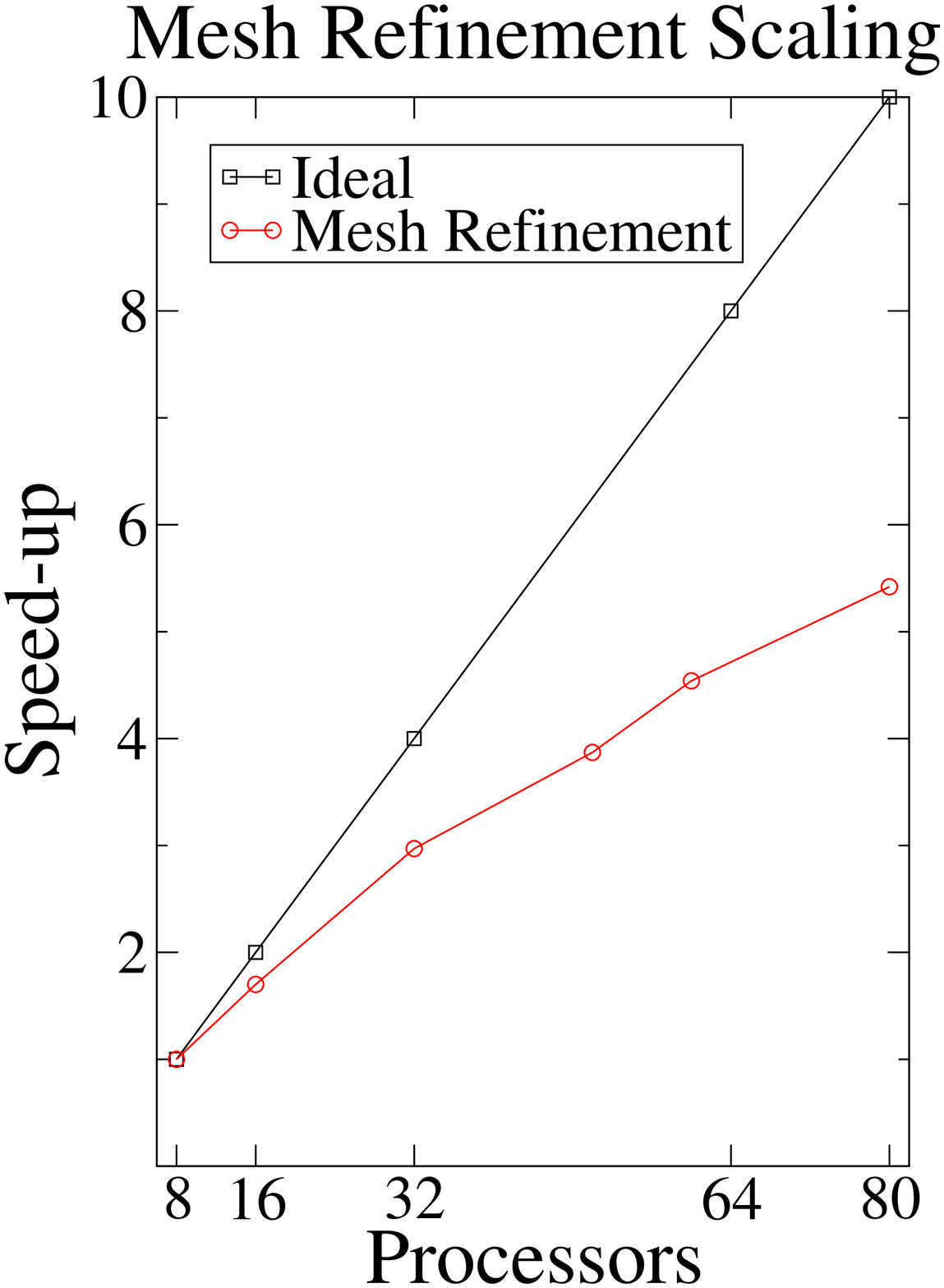,height=7.5cm}&\epsfig{file=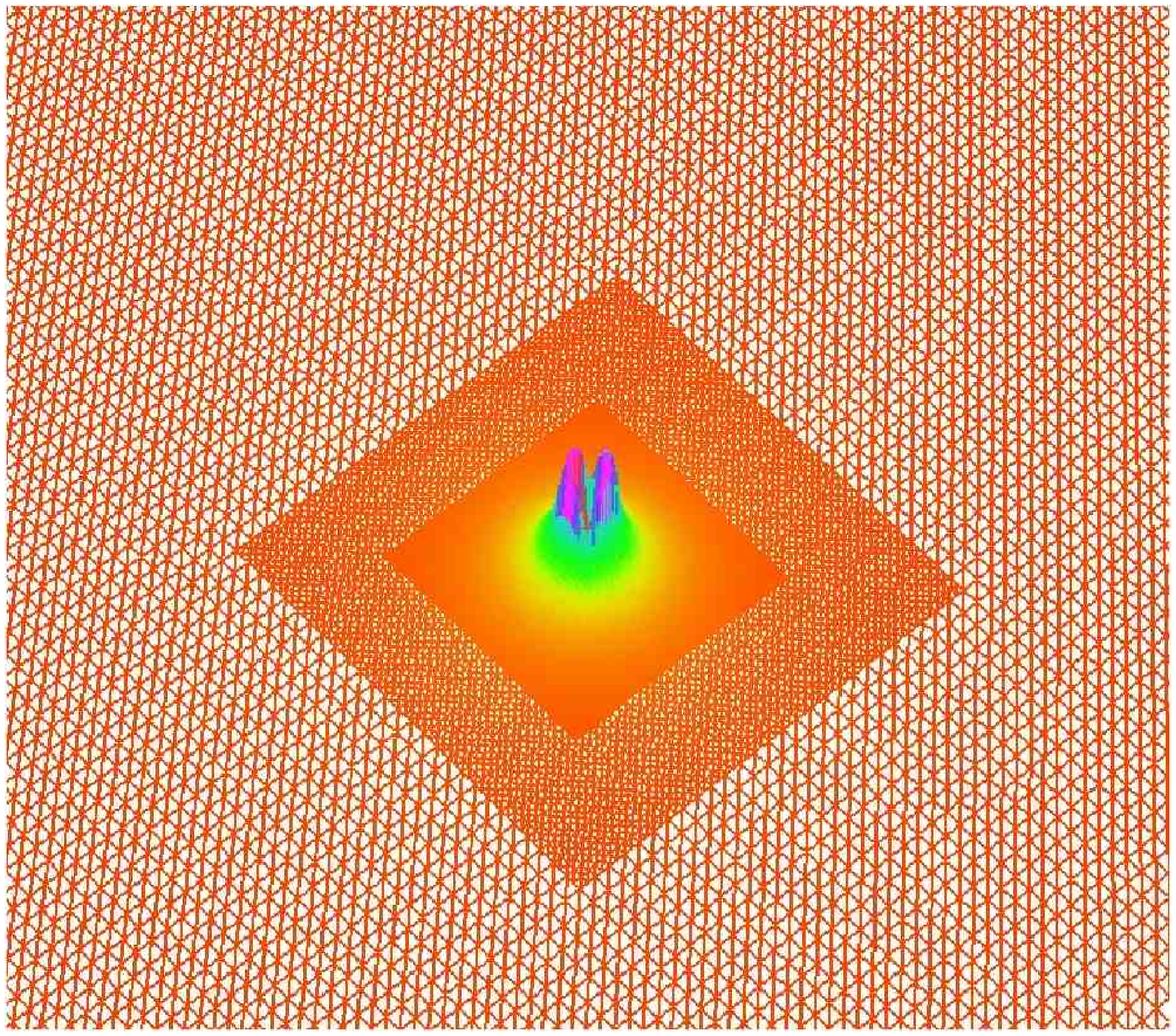,height=5.8cm}
\end{tabular}
\caption{
%This figure shows $\rho_0$ and the AMR grid structures at $t=0M$
%
% Should this be $\rho$ and not $\rho_0$?
%
The left frame shows a strong scaling test of our MHD code using
AMR.  In this test, thirty iterations were performed on a coarse grid of size
$81^3$ and a single level of refinement.
The right frame shows the rest density $\rho_0$ and the 
AMR grid structures for the Michel solution at $t=50M$ in the $x$-$y$ plane.
The domain of simulation is $\{x,y,z\}\in\left[-15M,15M\right]$.  The cubical
excision region is highlighted in the center of the grid on the
left.
}
\label{fig:accretion2}
\end{figure}

Astrophysical simulations of compact objects require that a large number
of processors can be used efficiently.  A rigorous measure of such
performance is the strong scaling test, where a model problem of fixed size
is run on increasing numbers of processors.
These tests indicate that the GRMHD code uses the distributed parallel
computing environment relatively efficiently, as our code scales
approximately linearly as the number of processors is increased by a factor
of ten~\cite{Anderson:2006}.  (See Figure~\ref{fig:accretion2}.)
Note, since the problem size is fixed, the
scaling can not be linear indefinitely.

To verify that the equations are implemented correctly, we have
compared results with exact solutions, and we discuss two of these tests
here: the Michel solution, and Tolman--Oppenheimer--Volkoff (TOV) 
solutions.
The Michel solution describes the continuous spherical accretion of a 
fluid onto a Schwarzschild black hole in the presence of a radial magnetic
field.  We use in-going
Eddington-Finkelstein coordinates in our calculation, and excise
a centrally located cubical region of  half width $0.3~M$ to remove the
singularity.  In this test, the fluid is
initially set to the Michel solution for radius $r > 2.5M$ while for $r \leq 2.5 M$ a constant pressure and density are chosen.  The system is
then evolved until a steady state is reached.  The refinement criterion
is based on the estimation of the truncation error provided by the shadow hierarchy.  
Figure~\ref{fig:accretion2} shows the AMR grid structures at time
$t=50M$.

\begin{figure}
\begin{center}
\epsfig{file=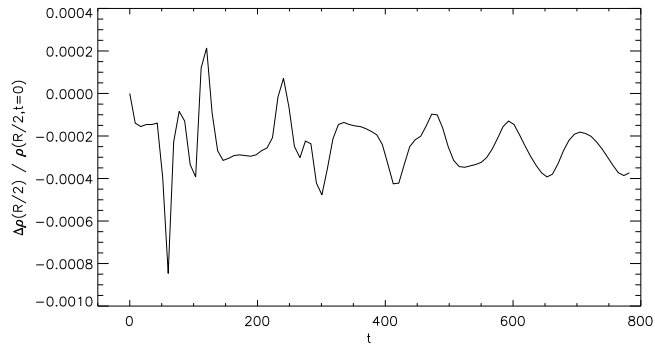,height=5.0cm}
\caption{
%This figure shows the variation in the density at $R/2$ for a stable TOV 
The variation in the density at $R/2$ for a stable TOV 
solution.  The evolution is performed
using $129^3$ points on a cubical domain  $\{x,y,z\}\in\left[-11M,11M\right]$.
The central density is $8.1\times 10^{-4}$, the stellar
radius $R=9.279$, and mass $M=0.5659$.
}
\label{fig:tov}
\end{center}
\end{figure}

A stable TOV solution is used for our second test,  
which we evolve in the Cowling approximation (fixed geometry) for over 400
light-crossing times.  The star oscillates as expected, and the oscillations
of the density
at half the stellar radius ($R/2$) are shown in Figure~\ref{fig:tov}.
The equation of state for the initial data is $P=\kappa\rho_0^\Gamma$, 
with $\Gamma =5/3$ and $\kappa = 4.349$.
Similar runs on dynamic backgrounds have also been performed
and show similar results, though for slightly shorter periods of time.

\section*{Acknowledgments}
We are pleased to thank Luis Lehner, Patrick Motl, and Ignacio Olabarrieta
for numerous suggestions and discussions during the course of this work.
We also thank Bruno Giacomazzo, Carlos Palenzuela, 
Oscar Reula, Luciano Rezzolla, and Joel Tohline for helpful 
discussions.
This work was supported by the NSF through the grants PHY-0326311, 
PHY-0244699, PHY-0326378, PHY-0502218, and PHY-0325224.

\bibliographystyle{ws-procs975x65}
\bibliography{main}

\end{document}